\documentclass[12pt,preprint]{aastex}

\def\beq{\begin{eqnarray}}
\def\eeq{\end{eqnarray}}

\begin{document}

\title{CONSTRAINTS ON THE MASS ACCRETION RATE OF NEUTRINO-COOLED DISKS IN GAMMA-RAY BURSTS}

\author{Tong Liu, Wei-Min Gu, Li Xue, Shan-Shan Weng, and Ju-Fu Lu*}

\affil{Department of Physics
and Institute of Theoretical Physics and Astrophysics, \\
Xiamen University, Xiamen, Fujian 361005, China}

\email{*lujf@xmu.edu.cn}

\begin{abstract}
 We present a unified description of all the three known classes of optically
thick accretion disks around black holes, namely Shakura-Sunyaev
disks, slim disks, and neutrino-dominated accretion flows (NDAFs).
It is found that NDAFs have both a maximal and a minimal possible
mass accretion rate at their each radius. This may be suggestive of
an interpretation for the origin of X-ray flares observed in
gamma-ray bursts.
\end{abstract}


\section{Introduction}

Recently, Gu \& Lu (2007) addressed a theoretical problem regarding
the slim accretion disk model in the fundamental sense. In this
model (e.g., Abramowicz et al. 1988; Kato et al. 1998, p.242), in
dealing with the vertical hydrostatic equilibrium of the disk, the
well-known pseudo-Newtonian potential of Paczy\'nski \& Wiita
(1980), \beq\ \psi (r,z) = - \frac {G M}{\sqrt{r^2+z^2}-r_{\rm g}},\
\eeq was approximated in the form suggested by H\={o}shi (1977),
i.e.,\beq\ \psi (r,z) = \psi (r,0) + \Omega_{\rm K} ^2 z^2/2,\ \eeq
where $r$ and $z$ are cylindrical coordinates, $M$ is the mass of
the central black hole, $r_{\rm g} \equiv 2 G M / c^{2} $ is the
Schwarzschild radius, and $\Omega_{\rm K} = (GM /r)^{1/2} /(r -
r_{\rm g})$ is the Keplerian angular velocity. As shown by Gu \& Lu
(2007), equation (2) is valid only for geometrically thin disks such
as Shakura-Sunyaev disks (SSDs, Shakura \& Sunyaev 1973), i.e., with
$H \ll r$, where $H$ is the half-thickness of the disk; and this
equation is invalid for slim disks that may have $H \lesssim r$ or
$H \sim r$, because in this case it would greatly magnify the
gravitational force in the vertical direction of the disk, $\partial
\psi /
\partial z$. When the explicit form of Paczy\'nski-Wiita potential,
equation (1), is used to calculate the vertical gravitational force,
the relationship that is obtained with equation (2) and is
applicable only for geometrically thin disks, i.e., $c_{\rm s} /
\Omega_{\rm K} H = {\rm constant}$, where $c_{\rm s}$ is the sound
speed, does not hold for slim disks. Accordingly and more seriously,
it is found that slim disks cannot exist at large radii of black
hole accretion flows with large accretion rates, and only the inner
regions of these flows can possibly take the form of slim disks
provided accretion rates are effectively reduced by outflows from
the outer regions.

Slim disk are optically very thick in the vertical direction, such
that photons are sufficiently trapped within the disk and advected
along with the disk matter into the black hole. In this paper, we
extend the work of Gu \& Lu (2007) into another class of accretion
disks, namely neutrino-cooled accretion disks (or neutrino-dominated
accretion flows, NDAFs) around stellar-mass black holes. These disks
are known to be plausible candidates for the central engines of
gamma-ray bursts (Popham et al. 1999; Narayan et al. 2001; Kohri \&
Mineshige 2002; Di Matteo et al. 2002; Kohri et al. 2005; Lee et al.
2005; Gu et al. 2006; Chen \& Beloborodov 2007; Liu et al. 2007;
Kawanaka \& Mineshige 2007; Janiuk et al. 2007). They can be
regarded as an even more extreme case of optically thick disks,
since their density and temperature are so high ($\rho \sim
10^{10}$g cm$^{-3}$, $T \sim 10^{10} {\rm K}$) that photons are
completely trapped and only energetic neutrinos are emitted away.

\section{Equations}

NDAFs may have $H$ comparable to $r$ (e.g., Popham et al. 1999; Chen
\& Beloborodov 2007; Janiuk et al. 2007), so their hydrodynamics and
thermodynamics are expected to be similar to those of slim disks. We
write the continuity, radial momentum, angular momentum, and energy
equations of NDAFs in the formalism of Gu \& Lu (2007) for slim
disks, where equation (1) was used to integrate the vertical
hydrostatic equilibrium equation and the simple relation $c_{\rm s}
/ \Omega_{\rm K} H = {\rm constant}$ was abandoned: \beq {\dot M} =
-2 \pi r \Sigma v_{r} = {\rm constant} \ , \eeq \beq v_{r}\frac{{\rm
d} v_{r}}{{\rm d}{\rm ln} r} + c_{\rm s}^2 \frac{{\rm d}{\rm ln}
\Pi} {{\rm d}{\rm ln} r} - \Omega ^ {2} r^{2} = - \frac{\Omega_{\rm
K}^2 (r-r_{\rm g})^2}{\Sigma} \int_{-H}^{H}
\frac{\rho}{[\sqrt{1+(z/r)^2}-r_{\rm g}/r]^2 \sqrt{1+(z/r)^2}} {\rm
d} z \ , \eeq \beq {\dot M} (\Omega r^2-j_0) = 2 \pi \alpha r^2 \Pi
\ , \eeq \beq Q_{\rm vis} = Q_{\rm adv} + Q_{\nu} \ . \eeq  In these
equations, $\dot M$ is the mass accretion rate, $v_{r}$ is the
radial velocity, $\Omega$ is the angular velocity, $\Sigma =
\int_{-H}^{H} {\rho} {\rm d} z$ is the surface density, $\Pi =
\int_{-H}^{H} {p}  {\rm d} z$ is the vertically integrated pressure,
$\rho$ is the mass density, $p$ is the pressure, the sound speed is
defined as $c_{\rm s} = (\Pi/\Sigma)^{1 / 2}$, $j_0$ is an
integration constant representing the specific angular momentum
accreted by the black hole, and $\alpha$ is the Shakura-Sunyaev
viscosity parameter. The viscous heating rate is $Q_{\rm vis} =
{\dot M} \Omega ^2 f g / 2 \pi$, where $f = 1 - j/\Omega_{\rm K}
r^2$, $j = j_0/\omega$, $\omega = \Omega/\Omega_{\rm K}$ is assumed
to be a constant that is smaller than 1 (sub-Keplerian rotation),
and $g = - {\rm d ln} \Omega_{\rm K}/{\rm d ln} r$; the advective
cooling rate is $Q_{\rm adv} = \xi {\dot M} {c_{\rm s}}^2  / 2\pi
r^2$, with $\xi$ being a dimensionless quantity of the order of
unity (Kato et al. 1998, p.272). All these definitions and relations
are similar to those for slim disks. What is new in the case of
NDAFs, however, is that in the energy equation (6) there is a
cooling term due to neutrino radiation, $Q_{\nu}$, instead of the
cooling term due to photon radiation, $Q_{\rm rad}$, because of the
photon trapping. The neutrino cooling is expressed by a bridging
formula that is valid in both the neutrino optically thin and thick
regimes:\beq\ Q_{\nu}=\sum_{i} \frac{(7/8) {\sigma} T^4 }
{(3/4)[\tau_{{\nu}_i}/2+1/ \sqrt{3}+1/(3 \tau_{a,{\nu}_i})]}  \eeq
(Korhi et al. 2005), where $T$ is the temperature, $\tau_{{\nu}_i}$
is the total optical depth for neutrinos, $\tau_{a,{\nu}_i}$ is the
absorption optical depth for neutrinos, and the subscript $i$ runs
over the three species of neutrinos $\nu_{\rm e}$ , $\nu_\mu$ , and
$\nu_\tau$ (see, e.g., Kohri et al. [2005] and Liu et al. [2007] for
detailed analyses and calculations of these optical depths).
Accordingly, the equation of state is also different from that for
slim disks, it is written as \beq p = p_{\rm gas} + p_{\rm rad} +
p_{\rm e} + p_{\nu} ,\eeq where $p_{\rm gas}$ , $p_{\rm rad}$ ,
$p_{\rm e}$ , and $p_\nu$ are the gas pressure from nucleons,
radiation pressure of photons, degeneracy pressure of electrons, and
radiation pressure of neutrinos, respectively. Of these four
pressure components, there are only the first two for slim disks,
while the last two are newly appeared for NDAFs. Detailed
expressions for these four components are also given in, e.g., Kohri
et al. (2005) and Liu et al. (2007). An additional note about the
energy equation (6) is that, for simplicity, we ignore another
cooling term due to photodisintegration of $\alpha$-particles and
other heavier nuclei; in other words, we assume that in NDAFs all
heavy nuclei are already disintegrated into nucleons.

\section{Unified Description of SSDs, Slim Disks, and NDAFs}

With the procedure similar to that of Gu \& Lu (2007), thermal
equilibrium solutions at a certain radius can be obtained from
equations (3) - (6) and (8), with given constant parameters $M$,
$\dot M$, $\alpha$, and $j$. In our calculations we take $M=3
M_\odot$, $\alpha=0.1$, and $j=1.83 c r_{\rm g}$.

  As our main result, Figure 1 shows thermal equilibria of NDAFs
at each radius $r$ with corresponding accretion rates $\dot M$. The
left and right vertical axes are for $\dot M$ in units of the
Eddington accretion rate ${\dot M}_{\rm Edd} = 64 \pi GM / c
\kappa_{\rm es}$, where $\kappa_{\rm es} = 0.34 {\rm cm}^2 {\rm
g}^{-1}$ is the electron scattering opacity, and of $M_\odot$
s$^{-1}$ , respectively. To have a complete picture of all the known
classes of optically thick accretion disks around black holes, the
results of Gu \& Lu (2007) for SSDs and slim disks are also included
in the figure. It is seen that the $\dot M$-$r$ plane is divided
into four regions by five lines $a$, $b$, $c$, $d$, and $e$. As
described in Gu \& Lu (2007), the region below line $a$ is for
stable, photon radiation-cooled and gas pressure-supported SSDs; the
region between lines $a$ and $b$ is for unstable, photon radiation
pressure-supported but not yet advective cooling-dominated SSDs; the
region between lines $b$ and $c$ is for stable, advective
cooling-dominated and photon radiation pressure-supported slim
disks; and line $c$ represents a maximal possible accretion rate for
each radius, above which no thermal equilibrium solutions exist
because the viscous heating $Q_{\rm vis}$ is always larger than the
total cooling $(Q_{\rm adv} + Q_{\rm rad})$. What is new in the
figure is lines $d$ and $e$. The 'no solution' region above line $c$
extends upward till line $d$, above which the cooling due to
neutrino radiation, $Q_\nu$, becomes important or even dominant,
such that thermal equilibrium solutions can be established again,
and these are exactly NDAF solutions. Thus, line $d$ represents the
lower limit of $\dot M$ needed for NDAFs to be realized. Similar to
line $c$, line $e$ shows the upper limit of $\dot M$ for NDAFs,
beyond which the heating $Q_{\rm vis}$ is always too large to be
balanced by the total cooling $(Q_{\rm adv} + Q_{\nu})$, and no
thermal equilibrium solutions can exist. Therefore, the joining
point of lines $d$ and $e$ at $r \approx 185$$r_{\rm g}$, marked by
a filled circle, defines the maximal possible outer boundary of an
NDAF. We repeat that the finding of a radius-dependent maximal
possible accretion rate, ${\dot M}_ {\rm max}(r)$, for slim disks as
well as for NDAFs (i.e., lines $c$ and $e$) is because of the usage
of equation (1). The physical reason for the existence of ${\dot M}_
{\rm max}(r)$ is that the black hole's gravitational force in the
vertical direction, correctly calculated from equation (1), can only
gather some limited amount of accreted matter; beyond this limit the
pressure force would be too large to be balanced by the
gravitational force, and outflows are likely to be produced as a
result.

Note that in Figure 1 the upper limit line $c$ for slim disks and
the lower limit line $d$ for NDAFs join into a single line at $r
\thickapprox 18.6 r_{\rm g}$, and the region for slim disk solutions
and that for NDAF solutions combine into a single region, ranging
over several orders of magnitude in $\dot M$. There is no boundary
separating these two seemingly very different classes of black hole
accretion disks, slim disks and NDAFs. Both of them are optically
very thick for photons. As the accretion rate increases, the
processes of neutrino emission operate and become important, and the
accretion flow changes from the slim disk form to the NDAF form. By
also including SSD solutions, which are separated from slim disk
solutions by the unstable region and can possibly connect slim disk
solutions via limit-cycle oscillations (e.g., Szuszkiewicz \& Miller
2001; Li et al. 2007), Figure 1 does provide a unified description
of all the three known classes of optically thick accretion disks
around black holes.

Line $e$ in Figure 1 looks to indicate that for very small radii,
the allowed accretion rate of NDAFs could be as high as $\sim 10^5 -
10^6 M_\odot {\rm s}^{-1}$. One might worry that the density of the
flow matter there could reach up to the same as or even larger than
the nuclear density, then there would be uncertainties in
calculating the total pressure, the neutrino cooling rate and so on,
because such calculations need information about the nuclear
equation of state in detail which is highly model-dependent and
unclear at present. Our answer to this problem is that such a high
accretion rate for a small radius is only a theoretical upper limit
and is unlikely to be astrophysically realizable. All the related
theoretical models and numerical calculations have agreed that the
accretion rate needed to power a GRB is in the range of $\sim 0.01 -
10 M_\odot {\rm s}^{-1}$ (e.g., Popham et al. 1999; Kawanaka \&
Mineshige 2007). We plot in Figure 2 the mass density on the
equatorial plane, $\rho_0$, as functions of $r$ for four different
values of $\dot{M}$: $0.01$, $0.1$, $1$, and $10 M_\odot {\rm
s}^{-1}$. It is seen that even for $\dot{M}= 10 M_\odot {\rm
s}^{-1}$, $\rho_0$ is far below the nuclear density which is $\sim
10^{14}$$ {\rm g}$ ${\rm cm}^{-3}$. Our calculations also give that
(not drawn in Figure 2) only for an (unrealistically) high accretion
rate $\dot{M} \sim 10^6 M_\odot {\rm s}^{-1}$, $\rho_0$ could reach
up to $\sim 10^{14}$$ {\rm g}$ ${\rm cm}^{-3}$ in a very small
region $r \lesssim 4r_{\rm g}$. Thus, Figure 1 does not seem to
suffer the density problem.

Figure 3 shows thermal equilibrium solutions in the $\dot
M$-$\Sigma$ plane for fixed radii $r = 5r_{\rm g}$ (the solid line)
and $r = 100r_{\rm g}$ (the dashed line). Such a design is widely
found in the literature as it is a useful tool for the local
stability analysis: solutions on the lines with a positive
derivative $({\partial {\dot M}}/{\partial \Sigma}
> 0)$ are viscously stable, while those with a negative derivative
$({\partial {\dot M}}/{\partial \Sigma} < 0)$ are viscously unstable
(e.g., Kato et al. 1998). As seen from the figure, for a small
radius $(r = 5r_{\rm g})$ there is a continuous S-shaped sequence of
thermal equilibria, of which the lower and middle branches
correspond to stable SSDs and unstable SSDs, respectively; and the
upper branch extends from stable slim disk solutions to stable NDAF
solutions with increasing $\dot M$ and ends at the maximal possible
$\dot M$ for NDAFs (i.e., line $e$ in Fig. 1). For a large radius
$(r = 100r_{\rm g})$, however, the solution sequence is no more
continuous and is broken into two parts by the 'no solution' region
shown in Figure 1: the lower part is still for stable SSDs and
unstable SSDs, and the upper part is for stable NDAFs and is bounded
by the minimal and maximal possible $\dot M$(i.e., lines $d$ and $e$
in Fig. 1). Previous works (Kohri \& Mineshige 2002; Di Matteo et
al. 2002; Kawanaka \& Mineshige 2007) have proved that NDAFs are
stable. Our results here confirm this conclusion, but with an
additional remark that for NDAFs the allowed accretion rate has both
a lower and an upper limit.

All the above results are obtained with a one-dimensional (1-D)
analytical model, that is, as in most previous works on slim disks
as well as NDAFs (see the references in $\S 1$), we solve the
vertically integrated equations and do not consider the vertical
structure of the disk. Strictly speaking, simple 1-D models can be
valid only for geometrically thin disks such as SSDs. For slim disks
and NDAFs which are geometrically not thin, the multidimensional
effects on the disk's basic properties, e.g., the various timescales
and the cooling rates can be important or even crucial, so the
validity of 1-D approximations should be checked by at least
two-dimensional (2-D) studies. Recently, 2-D
radiation-hydrodynamical numerical simulations and analytical
treatments of supercritical accretion flows, to which slim disks and
NDAFs belong, have been made (Ohsuga et al. 2005; Ohsuga 2007; Kohri
et al. 2007). These works showed that it is indeed quite important
to consider carefully the vertical dependence of fluid quantities
and the radiative diffusion in the vertical direction. Compared with
the correct 2-D results, the 1-D slim disk model (Abramowicz et al.
1988; Kato et al. 1998) tends to overestimate $Q_{\rm rad}$, and
hence, underestimate $Q_{\rm adv}$ and overestimate the luminosity.
Another important multidimensional effect for slim disks is that
photon trapping modifies the spectral energy distribution. For
NDAFs, however, similar multidimensional effects were only predicted
but not proved (Kohri et al. 2007). Having mentioned these recent
2-D works, we think that the validity of our 1-D results here is
unlikely to be affected. We do not calculate the luminosity or the
radiation spectrum. What we have done is a dynamical study. We pay
attention to the thermal equilibrium between the viscous heating
$Q_{\rm vis}$ and the total cooling, which is the sum of $Q_{\rm
rad}$ and $Q_{\rm adv}$ for slim disks and the sum of $Q_{\nu}$ and
$Q_{\rm adv}$ for NDAFs. Even though $Q_{\rm rad}$ (and maybe
$Q_{\nu}$ too) may be overestimated and $Q_{\rm adv}$ underestimated
in the 1-D calculations, these two inaccuracies should tend to
cancel with each other, such that the resulting total cooling is
likely to remain a reasonable estimate. From the analysis of thermal
equilibrium solutions we find, as our main result, that NDAFs have
both a maximal and a minimal possible accretion rate at their each
radius. This result implies that outflows driven by the pressure
force are likely to originate from accretion flows with an accretion
rate exceeding its allowed upper limit. Such outflows have been
found in the 2-D simulations of supercritical accretion flows
(Ohsuga 2007).

\section{On the Origin of X-Ray Flares in Gamma-Ray Bursts}

We discuss briefly an astrophysical implication of our results. A
new, unexpected phenomenology of gamma-ray bursts (GRBs) revealed by
the $Swift$ satellite is erratic X-ray flares following a number of
long-duration and short-duration GRBs. Such flares generally rise
and fall rapidly, with typical time scales much shorter than the
epoch when the flare occurs. This time behavior strongly supports
the 'internal' origin of the flares, in contrast to the 'external'
origin of the power-law decay afterglows; that is, it indicates a
restart of the GRB central engine (see, e.g., Proga \& Zhang [2006]
for references). Although long and short GRBs are likely to be
associated with different types of progenitors, namely collapsars
for long GRBs and mergers of compact objects for short ones; it is
generally believed that the central engines of these two classes of
events are in common invoking a hyperaccretion disk around a
stellar-mass black hole (see, e.g., Perna et al. [2006] for
references). Thus, a restart of the GRB central engine is just a
restart of accretion. Based on these considerations, models for a
common origin of the X-ray flares in both long and short GRBs have
been recently proposed, in which accretion of a black hole is
repeatedly ended and then restarted either by the fragmentation in
the outer parts of accretion flows (Perna et al. 2006), or by the
magnetic flux accumulated in the vicinity of the black hole (Proga
\& Zhang 2006).

  Taking the view that hyperaccretion disks radiate their gravitational
energy as neutrinos to power GRBs (i.e., NDAFs, see references
listed in \S 1), our results here may be suggestive of a new and
simpler interpretation for the cause of ending and restarting the
accretion process. As seen from Figure 1, NDAF solutions can exist
only for the region bounded by lines $d$ and $e$, the joining point
(the filled circle) of these two limit lines defines the largest
possible outer boundary, $r^{\ast} \approx 185 r_{\rm g}$, of an
NDAF and corresponds to an accretion rate ${\dot M}^{\ast} \approx
0.62 M_{\sun} {\rm s}^{-1}$. Thus, depending on the place where an
inflow toward the black hole originally starts and the available
mass supply rate, there are following three distinct cases. (1) The
inflow starts at a radius $r < r^{\ast}$ and the mass supply rate is
within the region between lines $d$ and $e$. In this case, an NDAF
forms and a GRB appears, but no flares follow. (2) The inflow either
starts at $r < r^{\ast}$ with a mass supply rate below line $d$, or
starts at $r > r^{\ast}$ with a mass supply rate less than ${\dot
M}^{\ast}$. This is the case in which no NDAFs can form and no GRBs
are produced at all. (3) The inflow either starts at $r < r^{\ast}$
with a mass supply rate above line $e$, or starts at $r > r^{\ast}$
with a mass supply rate larger than ${\dot M}^{\ast}$. In this case,
the flow must first lose matter in the form of outflows (because the
pressure force exceeds the gravitational force, as mentioned in \S
3), such that its remaining matter ensures an accretion rate within
the region between lines $d$ and $e$, then there is an NDAF and a
GRB. This primary hyperaccretion ends when the matter in the NDAF is
all swallowed by the central black hole as in case 1, but this is
not all. Some or all of the outflowing matter can be attracted
backward again by the black hole. It is this feedback that is
responsible for restarting accretion. Such secondary accretion is
expected to have a rate that is substantially lower than the primary
hyperaccretion rate and is insufficient to support another NDAF, so
it is likely to be in the form of photon radiation-cooled disks that
produce X-ray flares. In this line, a detailed model for X-ray
flares in GRBs is being worked out.

For the moment, let us add a few more words about the realistic
initial conditions of hyperaccretion flows and the possibilities of
the mechanism for X-ray flare production suggested here, by
comparing the results for $\dot{M}$ and the outer boundary of NDAFs
given in Figure $1$ with those existed in the literature. As already
mentioned in $\S 3$, in all the known theoretical models and
numerical calculations $\dot{M}$ needed to power a GRB is in the
range of $\sim 0.01  -  10 M_\odot {\rm s}^{-1}$. Further, the most
probable $\dot{M}$ for NDAFs is in the range of $\sim 0.2  -  1
 M_\odot {\rm s}^{-1}$ (Kawanaka \& Mineshige 2007). For the outer
boundary of NDAFs, defined as the radius inside which neutrino
cooling is significant, Chen \& Beloborodov (2007) calculated that
it is $\sim 100 r_{\rm g}$ for $\alpha = 0.1$ and $\sim 200 r_{\rm
g}$ for $\alpha = 0.01$. All these values are consistent with the
NDAF region bounded by lines $d$ and $e$ in Figure $1$. Regarding
the realistic conditions in the progenitor systems of GRBs, it is
less clear where the inflow originally starts and how large the
initial mass supply rate is, but it seems plausible that short GRBs
resulted from mergers of compact objects have smaller accretion
disks and long GRBs resulted from collapsars have larger ones (e.g.,
Popham et al. 1999; Narayan et al. 2001). In view of these facts,
there should be possibilities for case $3$ described above to be
realized. For instance, if an inflow starts at $r < r^*$ with a mass
supply rate above line $e$, then the resulting GRB is likely a short
one with flares following it; or if an inflow starts at $r > r^*$
with a mass supply rate larger than ${\dot{M}}^*$, then it is the
chance for a long GRB instead, and with flares too.

\acknowledgments

We thank Z. G. Dai for beneficial discussions and the referee for
helpful comments. This work was supported by the National Natural
Science Foundation of China under grants 10503003 and 10673009 and
the Natural Science Foundation of Fujian Province of China under
grant V0750001.

\clearpage

\begin{figure}
\plotone{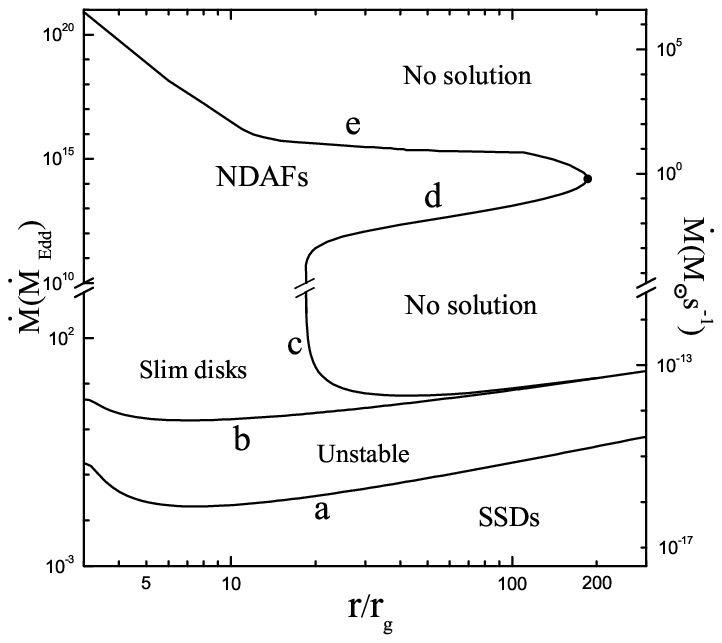} \caption{ Distribution of black hole optically
thick accretion disk solutions. The $\dot{M}$-$r$ plane is divided
into four regions by five boundary lines $a$, $b$, $c$, $d$, and
$e$. The filled circle denotes the largest possible outer boundary
of an NDAF. \label{fig1}}
\end{figure}
\clearpage

\begin{figure}
\plotone{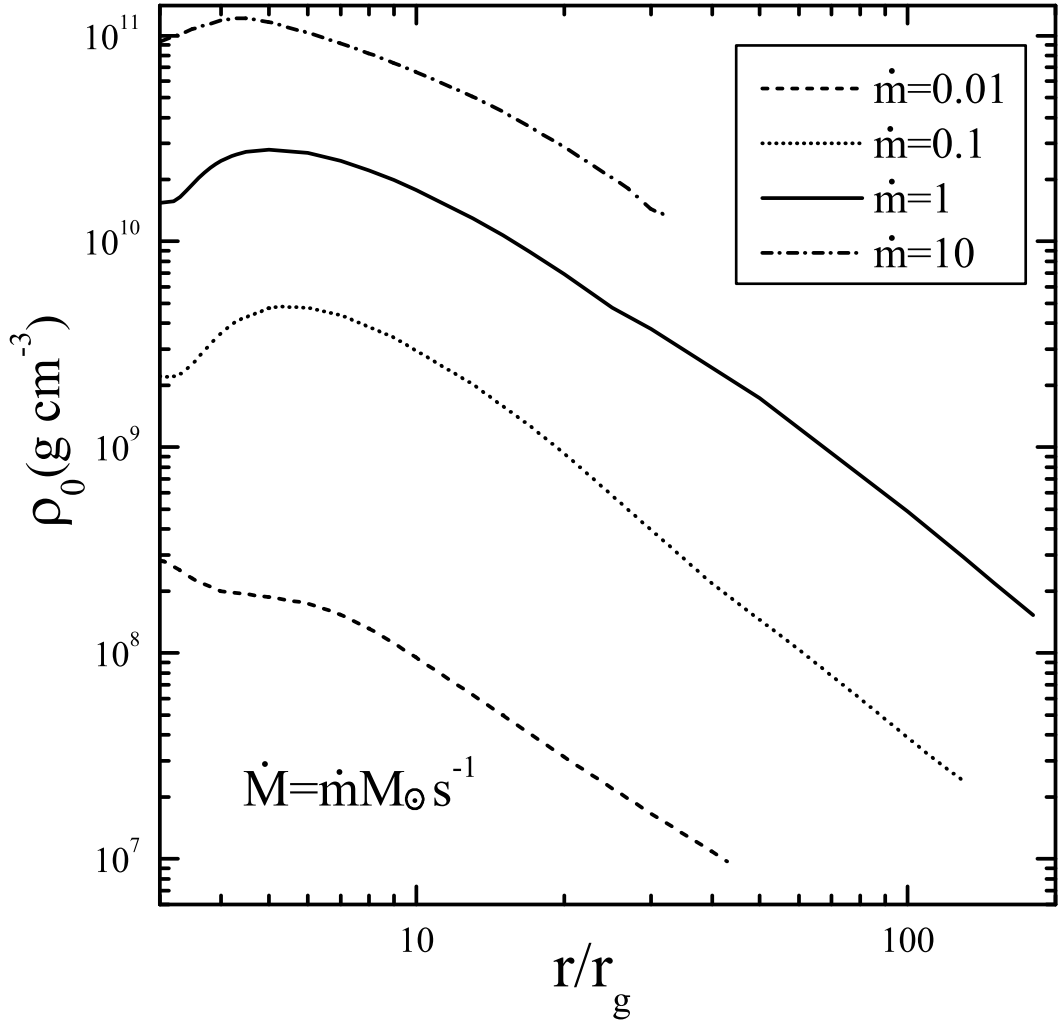} \caption{Radial dependence of the mass density
$\rho_0$ of NDAFs for different possible values of $\dot{M}$: $0.01$
(dashed line), $0.1$ (dotted line), $1$ (solid line), and $10$
(dot-dashed line) $M_\odot {\rm s}^{-1}$. \label{fig2}}
\end{figure}
\clearpage

\begin{figure}
\plotone{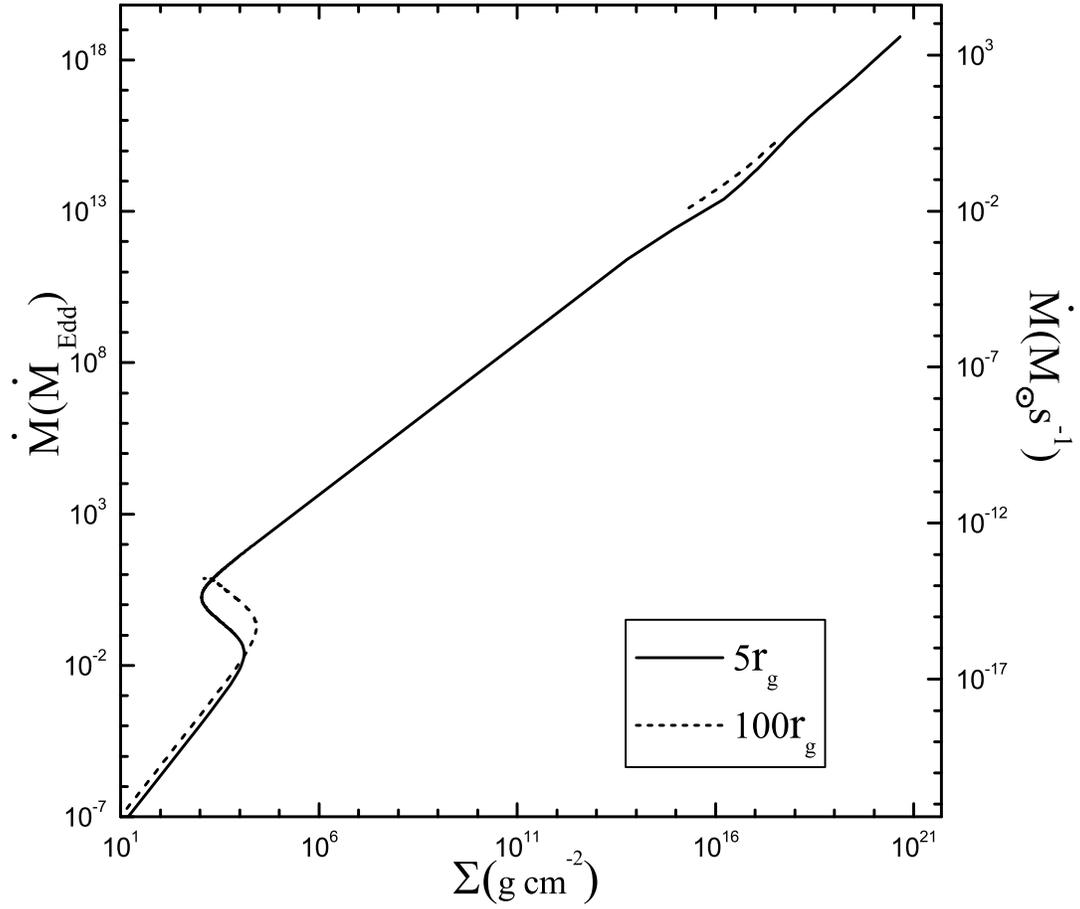} \caption{Stability curves in the $\dot{M}$-$\Sigma$
plane for radii $5 r_{\rm g}$ (solid line) and $100 r_{\rm g}$
(dashed line). \label{fig3}}
\end{figure}

\clearpage

\end{document}